\documentstyle[floats,prl,aps]{revtex}
\begin{document} 
\pagestyle{plain}
\setlength{\textwidth}{16cm}
\setlength{\textheight}{21cm}
\setlength{\headheight}{0cm}
\setlength{\topmargin}{0cm}
\setlength{\oddsidemargin}{0cm}
\setlength{\parskip}{0.5cm}
\newcommand{\be}{\begin{equation}}
\newcommand{\ee}{\end{equation}}
\newcommand{\beqa}{\begin{eqnarray}}
\newcommand{\eeqa}{\end{eqnarray}}
\draft 
\title{The Universe as a Diffusive Medium -- Constraining or Detecting the 
Gravitational Wave Background}
\author{Francine R. Marleau$^1$ and Glenn D. Starkman$^2$} 
\address{$^1$ Department of Astronomy, Campbell Hall, University of California, 
Berkeley, CA 94720}
\address{$^2$Department of Physics, Case Western Reserve University,
Cleveland, OH 44106-7079} 
\date{\today}
\maketitle
\begin{abstract}
We calculate the ``seeing'' effect on distant sources due to a gravitational 
wave background.  We derive the limit in strain and energy density of the 
gravitational wave based on the limit of detectability of 
this effect with the present day telescope resolution.  
We also compare our detection limit to those obtained from existing methods.
\end{abstract}

%\keywords{gravitational waves}

\section{Introduction}

The generation of gravitational waves is believed to be
an ongoing process in the evolution of the universe. 
Presently, aspherical supernovae and the merger of compact binaries 
are probably the two most common (or at least prosaic)
important sources  of gravitational waves.
Gravity waves may also have been copiously  produced in the early universe
and would probably have led to a nearly homogeneous and isotropic
background of gravity waves.
The most well-motivated model predicting a gravitational wave background 
from the early universe is inflation (Grishchuk; Ford and Parker;
Starobinsky; Rubakov, Sahin and Veryaskin; Fabbri and Pollock; Abbot and Wise;
Starobinsky; Abbot and Shaefer; Abbot and Harari; Allen;
Ressel and Turner; Sahni;  Souradeep and Sahni; Liddle and Lyth;
Davis et al.; Salopek; Lucchin, Matarrese and Mollerach;
Dolgov and Silk; Turner; Crittenden et. al.; Harari and 
Zaldarriaga; Crittenden, Davis and Steinhardt; Ng and Ng; Krauss \& White;
White; White, Krauss and Silk; Bond et. al.; Grischuk; Falk, Rangaranjan
and Srednicki; Luo and Schramm; Srednicki).  

The effect of this gravitational wave background on pulsar timing measurements 
has already been investigated (Bertotti {\it et al.} 1983).  
It has been also been studied in the context of the   
cosmic microwave background radiation when computing the 
Sachs-Wolfe contribution (Krauss \& White 1992; Davis {\it et al.} 1992) 
and the polarization of the 
radiation (Polnarev 1985; Crittenden {\it et al.} 1993).  
Fakir (1993) has shown how individual gravity waves bend lightlike geodesics.
For distant objects such as quasars, 
one therefore expects the gravitational wave background 
to perturb the light coming to us  and distort the image,
creating a ``seeing'' effect.  
The detection (or non-detection) of this effect provides us with an 
important additional constraint on models predicting 
the existence of a gravitational wave background.   

We present in the following sections a calculation of 
the expected deviation of light rays due to a gravitational wave background.  
We begin by recalculating the null geodesic deviation 
due to a single gravitational wave.  
We proceed to derive the expected RMS deviation,
with the deviation modeled as a random walk in three-dimensional space.  
We use that to put an upper limit on the dimensionless strain, $h$, 
and on the ratio, $\Omega_g$, of energy density in the gravitational
wave background to the critical energy density required to close the 
universe.  
This calculation is valid only if the dimension of the source 
is larger than the wavelength of the gravitational wave.  

\section{Deviation of Light by a Gravitational Wave} 

For a gravitational wave propagating in a Minkowski background spacetime, 
the metric can be written as:
\be
g_{\mu\nu} = \eta_{\mu\nu} + h_{\mu\nu}, 
\ee
where $\eta_{\mu\nu}$ is the Minkowski metric and $h_{\mu\nu}$ is 
the perturbation.  
The plane wave solution to Einstein's equations can then be written as:
\be
h_{\mu \nu}(x) = h \left(e_{\mu \nu} exp\left[ik_{\lambda}x^{\lambda}\right] + 
e^{\ast}_{\mu \nu} exp\left[-ik_{\lambda}x^{\lambda}\right]\right) \label{gw}
\ee
with $k$ null,
\be
k_{\mu}k^{\mu} = 0,
\ee
and
\be
k_{\mu} e^{\mu}_{\nu} = \frac{1}{2} k_{\nu} e^{\mu}_{\mu}\label{harmonic}
\ee
in harmonic gauge.
The polarization tensor $e_{\mu \nu}$ is symmetric, 
i.e. $e_{\mu \nu} = e_{\nu \mu}$.

For gravitational radiation travelling in the $-x$ direction 
\be
k^{\mu} = k(-1,-1,0,0).
\ee
The four-velocity of a null ray leaving the source in a general direction
is given by:
\be
u^{\mu} = (-1,\cos\theta,\sin\theta \cos\phi,\sin\theta \sin\phi).
\ee

The deviation of the null ray leaving the source is computed using 
the equations of parallel transport for the photon velocity
$u^\mu \equiv dx^\mu/d\lambda$, where $\lambda$ is the affine parameter 
along the null geodesic:
\be
\frac{du^{\mu}}{d\lambda} = 
- \Gamma^{\mu}_{\alpha \beta} u^{\alpha} u^{\beta}, \label{du} 
\ee
To leading order in $h$, 
\be
\Gamma^{\mu}_{\alpha \beta} = h \frac{i}{2} \eta^{\mu \nu} 
\{exp[ik_{\lambda}x^{\lambda}] (k_{\beta} e_{\nu \alpha} 
+ k_{\alpha} e_{\nu \beta} - k_{\nu} e_{\alpha \beta}) 
- exp[-ik_{\lambda}x^{\lambda}] (k_{\beta} e^{\ast}_{\nu \alpha} 
+ k_{\alpha} e^{\ast}_{\nu \beta} - k_{\nu} e^{\ast}_{\alpha \beta})\}.
\ee
Since the gravitational wave is propagating in the $-x$ direction,
equation (\ref{harmonic}) implies
\be
e_{ty}  = -e_{xy}, \; e_{tz} = -e_{xz}, \; e_{tx} = -(e_{xx} + e_{tt})/2, \; e_{yy}  = -e_{zz}. 
\ee
Taking the polarization to have only y and z components, 
the only non-vanishing Christoffels  are (up to
$\Gamma^{\mu}_{\alpha \beta} = \Gamma^{\mu}_{\beta \alpha}$)
\be 
\begin{array}{rl}
\Gamma^t_{yy} & =  -\Gamma^t_{zz} = \Gamma^y_{ty} = -\Gamma^z_{tz} 
= \frac{1}{2} h_{yy,t}, \\
- \Gamma^x_{yy} & =  \Gamma^x_{zz} = -\Gamma^z_{zx} = \Gamma^y_{yx} 
= \frac{1}{2} h_{yy,x}, \\
\Gamma^t_{yz} & =  \Gamma^y_{tz} =  \Gamma^z_{ty} = \frac{1}{2} h_{yz,t}, \\
- \Gamma^x_{yz} & =  \Gamma^y_{xz} =  \Gamma^z_{xy} = \frac{1}{2} h_{yz,x},
\end{array}
\ee

Therefore, the deviation of the null velocity vector is seen to be of ${\cal O}(h)$,
and is given by:
\be
\begin{array}{rl}
\frac{du^t}{d\lambda} &= \frac{1}{2} h_{yy,t} (u^{z2} - u^{y2}) , \\
\frac{du^x}{d\lambda} &= \frac{1}{2} h_{yy,x} (-u^{z2} + u^{y2}) , \\
\frac{du^y}{d\lambda} &= - (h_{yy,x} u^y u^x + h_{yy,t} u^t u^y 
+ h_{yz,t} u^t u^z + h_{yz,x} u^x u^z), \\
\frac{du^z}{d\lambda} &= (h_{yy,x} u^z u^x + h_{yy,t} u^t u^z 
- h_{yz,t} u^t u^y - h_{yz,x} u^x u^y) .
\end{array}
\ee
For null rays parallel or antiparallel to to the gravitional wave, 
this implies that the ray will not be subject to any deviation, as expected.  
Maximum deviation occurs when the incidence angle between 
the null ray and the gravitational wave is of $\pi/2$.  
The null vector in this case is $(-1,0,0,1)$ and:
\be
\begin{array}{rl}
\frac{du^t}{d\lambda} &= \frac{1}{2} h_{yy,t}  , \;\\
\frac{du^x}{d\lambda} &= -\frac{1}{2} h_{yy,x} , \;\\
\frac{du^y}{d\lambda} &= h_{yz,t} , \;\\
\frac{du^z}{d\lambda} &= -h_{yy,t}  .\\
\end{array}
\ee
As the photon travels through  a wave train, its direction oscillates with
angular amplitude $\alpha\simeq h$.

\section{Total Deviation of Light Ray by Random Walk Process} 

The deviation of the null geodesic due to a single gravitational wave 
can be viewed as a single step in  a  random walk in three dimension.  
It is therefore possible to 
evaluate the magnitude of the expected deviation of the light of 
a distant source 
caused by a background of gravitational waves.  The random walk 
gives us the following relation:
\be
<\Delta \theta_{tot}> = \sqrt{\frac{N}{3}} \bar{\alpha}, 
\ee
where $\bar{\alpha} \sim h$ from the calculation above.  
$N$ is the number of average-size gravity wavetrains 
through which the light has passed and equals the ratio of the
source-observer distance, $d_s$,
 to the average coherence length of the gravity wave background, $\ell_{coh}$.  
i.e.~$N = d_s/\ell_{coh}$.  The total deviation can be written as:
\be 
\Delta \theta_{tot} = 10^{-12} \; \frac{1}{\sqrt{\omega \ell_{coh}}}
\; \sqrt{\frac{\omega d_s}{300 \; Mpc \; Hz}} \; (\frac{h}{10^{-20}}),
\ee
where $\omega$ for primordial gravitational waves ranges from 
the scale of the present horizon ($\sim 6000$ Mpc) down to 
microphysical scales.  
In this case, we have assumed that the average perturbation 
was of order $h$, which is independent of frequency.  Returning 
to the original perturbation equations, one sees that in fact they 
are frequency dependent since $h_{,x}$ and $h_{,t}$ depend on the 
wave number, i.e. $h = h(w)$.  

The ratio of the energy density 
in the gravitational wave background at a given frequency, $\omega$,
to the critical density for closure $\rho_c = 1.054 \times 10^{-5} \; h_0^2$ GeV/cm$^3$, 
is 
\be
\Omega_g(\omega) \equiv {1\over \rho_c} {d\rho \over d\log{\omega}}
= 3.17\times 10^{-6} \; h_0^{-2} \; \left(\frac{\omega}{Hz}\right)^2 \; 
\left(\frac{h}{10^{-20}}\right)^2.
\ee
This can be rewritten as
\be
h_{20}^2(\omega) \equiv \left(\frac{h}{10^{-20}} \right)^2
= 12 \; \Omega_g(\omega) \left({H_0\over 2\omega}\right)^2,
\ee
where $H_0 = 100 \; h_0$ km s$^{-1}$ Mpc$^{-1}$.  

\section{Detectability}

A photon emitted by a distant object and propagating along the line of sight
through the gravitational wave background, 
will be displaced in the sky compared to its position in the absence of the
background, by an angle of the order of $\Delta \theta_{tot}$. 
This displacement should vary as the gravitational
wave travels across space, creating a seeing disk for images of 
distant objects, such as distant galaxies or quasars.   
This angular deviation should be observable with high enough resolution.  
With the present high resolution telescopes such as the Hubble Space 
Telescope, it is easy to put an upper limit on the strength of primordial 
gravitational waves if the seeing effect is not observed.  
The Hubble Space Telescope has an angular resolution of the order of 
$\Delta \theta_{lim} \sim 1$ arcsec $\sim 5 \times 10^{-7}$ radians.  
This limit can be pushed down even more in the future with long 
bsim aseline interferometers ($\Delta \theta_{lim} \sim 10^{-3}$ arcsec $\sim 5 \times 10^{-9}$
radians) or even space-based interferometers ($\Delta \theta_{lim} \sim 10^{-7}$ arcsec 
$\sim 5 \times 10^{-13}$ radians) built to observe radio sources.   

Given that the seeing effect described above 
hasn't been seen with the present detection limit of $\Delta\theta_{lim}\simeq5\times10^{-7}$, 
we can infer an upper limit on the dimensionless
strain or energy density of the background gravitational wave 
from the condition $\Delta \theta_{tot} < \Delta\theta_{lim}$ which 
translates into:
\be
h < 2.5 \times 10^{-8} \; \Delta\theta_{lim} \; \sqrt{\ell_{coh}/\lambda} \; \sqrt{\frac{300 \; Mpc \; Hz}
{\omega d_s}},
\ee
and 
\be
\Omega_g \; h_0^2 < 1.98 \times 10^2 \; (\frac{\omega}{Hz}) \; (\ell_{coh}/\lambda) \;
(\frac{3 \; Gpc}{d_s})\left(\frac{\Delta\theta_{lim}}{10^{-8}}\right)^2
\ee
It is interesting to note that 
the dimensionless strain predicted by inflation (Bar-Kana 1994) ranges from 
$10^{-16}$ for a frequency, $\nu$, of $10^{-8}$ Hz to $10^{-28}$ for 
a frequency of $10^{4}$ Hz, assuming a strictly scale-invariant 
spectrum (no tensor fluctuations).  
The best limits currently quoted are from pulsar timing, 
which probes strains greater than $10^{-13}$ over a small 
range in frequency near $\sim 10^{-8}$ Hz.  The limit on 
the energy density of a gravitational wave background in this
frequency range is therefore $\Omega_g \; h_0^2 < 9 \times 10^{-8}$.  

The measurement of the quadrupole anisotropy produced by 
a gravitational wave of the cosmic microwave background has  
also been proposed  as 
a way of probing the strain of the background of 
gravitational wave although it is difficult to separate  
the observed signal caused by the density perturbations 
and the one due to the gravitational wave.  

Our method is an alternative astrophysical method and has the advantage 
of probing smaller strains and a wide range of frequencies.
Taking as our representative object a quasar at 3 Gpc $h_0^{-1}$,
and using $\Delta\theta_{tot}\leq 5\times10^{-7}$,
the new limits on $h$ range approximately from $3 \times 10^{-11}$ 
for a frequency, $\nu$, of $10^{-8}$ Hz to $3 \times 10^{-17}$ for a frequency of $10^{4}$ Hz.  
As seen in Fig.~\ref{f1}, it gives weaker limits than the pulsar timing method 
for the same small range of frequency 
but covers a much wider range of frequencies.
As minimum angular resolutions improve, 
with the development of long baseline inteferometers,
these limits could improve, by as much as $10^3$.
Ultimately, it might be possible to use this method 
to detect the background of gravitational waves.
  
\acknowledgments{
G.S. thanks Lawrence Krauss and Scott Koranda for very helpful discussions.
F.M. would like to thank the Physics Department at Case Western Reserve University for 
its hospitality and would like to acknowledge the support of Space Sciences Laboratory.}

\newpage

\begin{figure}
\caption{ Limits on the dimensionless strain.  The solid line 
is from gravitational seeing for an object 3 Gpc $h_0^{-1}$ distant and 
assuming $\ell_{coh}/\lambda = 1$.  The dotted line is 
the limit from pulsar timing.  The dashed line is 
a typical inflationary prediction (Bar-Kana 1994). } \label{f1}
\end{figure}


\begin{references}
\def\mnras{Mon. Not. Roy. Ast. Soc.}
\def\apjl{Astrophys. J. Lett.}
\def\prl{Phys. Rev. Lett}
\def\prd{Phys. Rev. D}
\def\sovast{Sov. Ast.}
\bibitem{Grisha} Grishchuk, L.P., Lett. Nuovo Cimento {\bf 12}, 60 (1975);
Sov. Phys. JETP  {\bf 40}, 409 (1975).
\bibitem{FordPark}  Ford, L.H. and  Parker, L., Phys. Rev. {\bf D16}, 245 
(1977);{\bf D16}, 1601 (1979).
\bibitem{Staroba} JETP Lett. {\bf 30}, 682 (1979).
\bibitem{Rubakov} Rubakov, V.A., Sazhin, M.V., and Veryaskin, A.V., Phys Lett 
{\bf 115B}, 189 (1982).
\bibitem{Fabbri} Frabbri, R. and Pollock, M.D., Phys Lett. {\bf 125B}, 445 
(1983).
\bibitem{AbbotWise} Abbot, L.F. and Wise, M.B., Nucl. Phys. {\bf B244}, 541
(1984); Phys. Lett. {\bf 135B}, 279 (1984).
\bibitem{Starobb} Sov. Astron. Lett. {\bf 9}, 302 (1983); {\bf 11}, 133 (1985).
\bibitem{AbbotShaefer} Abbot, L.F. and Shaefer, R.K., 
Astrophys. J. {\bf 308}, 546 (1986).
\bibitem{AbbotHarari} Abbot, L.F. and Harari, D.D., Nucl. Phys. 
{\bf 264}, 487 (1986).
\bibitem{Allen} Allen, B., \prd {\bf 37}, 2078 (1988).
\bibitem{Sahni} Sahni, V., \prd {\bf 42}, 453 (1990).
\bibitem{Souradeep} Souradeep, T. and  Sahni, V., Mod. PHys. Lett. A {\bf 7},
3541 (1992).
\bibitem{Liddle} Liddle, A.R. and Lyth, D.H., Phys. Lett. B 
{\bf 291}, 391 (1992).
\bibitem{Davis92} Davis, R.L., Hodges, H.M., Smoot, G.F., Steinhardt, P.J. and
Turner, M.S., \prl {\bf 69}, 1856 (1992). 
\bibitem{Salopek} Salopek, D.S., \prl {\bf 6}, 3602 (1992).
\bibitem{Lucchin}Lucchin, F., Matarrese, S. and Mollerach., S., 
Astrophys. {\bf 401}, L49 (1992).
\bibitem{Dolgov}Dolgov, A. and Silk, J., \prd {bf 47} 2619 (1993).
\bibitem{Turner}Turner, M.S., \prd {\bf 48}, 3502 (1993).
\bibitem{Crittenden92}Crittenden, R., Davis., R.L. and Steinhardt, P.J., \apj
{\bf 417}, L13 (1993).
\bibitem{Harari}Harari, D.D. and  Zaldarriaga, M., Phys. Lett. B {\bf 319},
96 (1993).
\bibitem{NgNg} Ng, K.L. and Ng, K.-W, Taiwan Inst. of Phys. Rep. No.
 IP-ASTP-08-93, astro-ph 9305001 (unpublished).
\bibitem{Krauss92} Krauss, L.M. and White, M., \prl 69, 869 (1992).
\bibitem{White} White, M. \prd {\bf 46}, 4198 (1992).
\bibitem{WKS} White, M., Krauss, L.M. and Silk, J., \apj {\bf 418}, 535 (1993).
\bibitem{Bond} Bond, J.R., Crittenden, R., Davis, R.L., Efstathiou, G. and 
Steinhardt, P.J., \prl {\bf 72}, 13 (1994).
\bibitem{Grishb}Grishchuk, L.P., \prl {\bf 70}, 2371 (1993).
\bibitem{Grishb}Grishchuk, L.P., Class. Quantum Grav. {\bf 10}, 2449 (1993);
\prd {\bf 48}, 5581 (1993); {\bf 46} 1440 (1992).
\bibitem{Falk} Falk, T., Rangarajan, R., and Srednicki, R., \apj {\bf 403}, L1 (1993).
\bibitem{Luo} Luo, X. and Schramm, D.N., \prl {\bf 71}, 1124 (1993).
\bibitem{Srednicki} Srednicki, M., \apj {\bf 416}, L1 (1993).
\bibitem{Fakirpre} Fakir, R., 1993 preprint, 
``Early Direct Detection of Gravity Waves.''
\bibitem{Fakir93} Fakir, R.,  \apj 418, 202 (1993).
\bibitem{Bertotti83} Bertotti, B., Carr, B.J., and 
Rees, M.J. 1983, \mnras 203, 945 
\bibitem{Polnarev85} Polnarev, A.G. 1985, \sovast 29, 607 
\bibitem{Bar-Kana94} Bar-Kana, Rennan 1994, \prd 50, 1157 
\end{references}
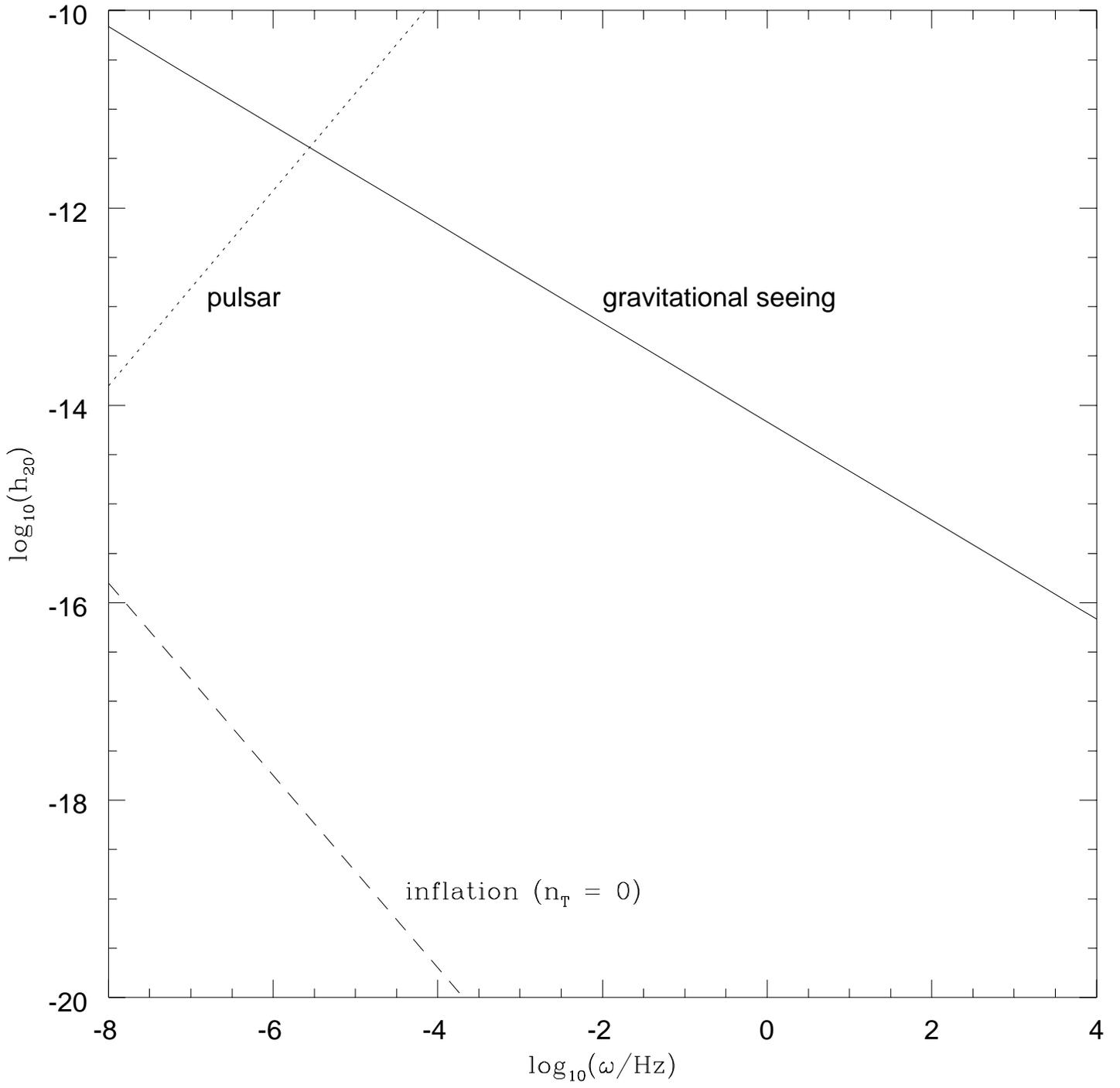
\end{document}